\documentclass[12pt,a4paper]{report}
\usepackage{setspace}
\def\ref {\par\noindent\hangindent 20pt}
\begin{document}
\def \bea {\begin{eqnarray}}
\def \eea {\end{eqnarray}}
\begin{center}
 {\large \bf{2D Monte-Carlo Radiative transfer modeling of the disk shaped secondary of Epsilon Aurigae }} \\
\vskip 1.3cm
{\bf{C. Muthumariappan$^{1}$ $\&$ M. Parthasarathy$^{2,3}$}} \\
\vskip .5cm
\end{center}
{\bf{ 1) Vainu Bappu Observatory, Indian Institute of Astrophysics \\
Kavalur, 635 701, India. e-mail: $muthu@iiap.res.in$  \\
2)  National Astronomical Observatory of Japan, 2-21-1 Osawa, Mitaka, Tokyo 181-8588, Japan \\
3) Aryabhatta Research Institute of Observational Sciences, Nainital, India }} \\
\vskip 1.2cm

\begin{center} 
{\bf {Abstract}} \\
\end{center} 
We present two dimensional Monte-Carlo radiative transfer models for the disk of the eclipsing binary $\epsilon$ Aur  by fitting its
spectral energy distribution from optical to the far-IR wavelengths. We also report new observations of $\epsilon$ Aur made by AKARI
in its five mid and far-IR photometric bands and were used to construct our SED. The disk is optically thick and has flared disk
geometry containing gas and dust with a gas to dust mass ratio of 100. We have taken the primary of the binary to be a F0Iae-type post-AGB
star and the disk is heated by a B5V hot star with a temperature of 15,000 K at the center of the disk. We take the radius of the disk
to be 3.8 AU for our models as constrained from the IR interferometric imaging observations of the eclipsing disk. Our models imply
that the disk contains grains which are much bigger than the ISM grains (grain sizes 10$\mu$ to 100$\mu$). The grain chemistry of the
disk is  carbonaceous and our models show that silicate and ISM dust chemistry do not reproduce the slope of the observed SED in the 
mid-IR to far-IR regions. This implies that the formation of the disk shaped secondary in $\epsilon$ Aur system could be the result of
accretion of matter and or mass transfer from the primary which is now a F0Iae post-AGB star. It is not a proto-planetary disk. The
disk is seen nearly edge on with an inclination angle larger than 85$^{o}$. We propose from our radiative transfer modeling that the
disk is not solid and  have a void of 2AU radius at the center within which no grains are present making the region nearly
transparent. The disk is not massive, its mass is derived to be less than 0.005M$_{\odot}$. \\

\begin{center}
{\bf {1. Introduction}}\\
\end{center}

\noindent
Epsilon Aurigae (HD 31964; $\epsilon$ Aur hereafter) is an eclipsing binary with an orbital period of 27.1 years showing
0.75 mag depth in optical during eclipse and the primary is occulted by the disk shaped secondary causing two year long eclipse
(Guinan $\&$ DeWarf 2002, Carroll et al. 1991, and Parthasarathy $\&$ Frueh 1986).
 The eclipse depth is independent of wavelength, but the depth, duration of the eclipse
and the masses of the components imply that the components should be almost equally bright, however,  no secondary eclipse was
observed. Eclipse characteristics indicate that the occulting object is very elongated with a dimension of 5-10 AU
parellal to the binary orbit.
 IR observations made by Woolf (1973) revealed an excess emission in the infrared. The observations carried out during the
previous eclpises and particulary the 1982 - 1984 eclipse and the recent 2009 - 2011 eclpse
 reveal the presence of a dusty plus gaseous disk in $\epsilon$  Aur, which is the body causing a two year long
eclipse. The presence of neutral gas in and around the disk shaped secondary was for the first time discovered
by Parthasarathy (1982) (see Stencel 1982), and Parthasarathy $\&$ Lambert (1983a) from the systematic increase in the strength of
KI 7699\AA~ line during the 1982 - 1984 eclipse.  \\
 
\noindent
 Orbital characteristics and spectral properties of the primary are consistent with two different models for the
system; one is a high mass star model with  the primary having a mass of 15M$_{\odot}$ and the other one is the low mass
star model with the primary having a mass of 4M$_{\odot}$. The optical spectra of $\epsilon$ Aur near the end of 1954-1956
eclipse was used to hypothesis the presence of a Be like hot star at the center of a large disk, from deducing the
electron density at the disk (Hack 1961). Later, the  IUE UV observations during the 1982 - 1984 eclipse
implied the presence of a hot source inside the disk which
can be fitted by a B5V star (Parthasarathy $\&$ Lambert (1983b),
 Ake $\&$ Simon (1984), Boehm, Ferluga $\&$ Hack (1984), Chapman, Kondo $\&$ Stencel (1983), Alter et al. (1986),
 Hoard et al. (2010)).\\

\noindent
 The mass ratio of the primary to the
secondary obtained from binary  solution is 0.62 (Kloppenborg et al., 2010) which constrains a lower mass for the the 
primary of $\epsilon$ Aur. Accurate distance measurement by Hipparcos (650pc, Perryman et al. 1997) and inturn an
estimate on intrinsic luminosity of the primary also does not favour for a high mass star model.  \\

\noindent
 Enhancement of N, Na and
$s$-process elements were found in the atmosphere of the primary by Sadakane et al.(2010) in their high resolution
optical spectroscopy. This may indicate the occurance of third dredge-up and $s$-process nucleosynthesis in the primary
of $\epsilon$ Aur, suggesting again a post-AGB primary for $\epsilon$ Aur. It is now fairly established  that the
$\epsilon$ Aur system consists a post-AGB star, a B5 type main sequence secondary and a disk of gas and  dust. \\

\noindent
    Recently Stefanik et al. (2010) reported an updated single-lined spectroscopic solution for the orbit of the
F0Iae primary star based on 20 years of monitoring at the CfA, combined with historical velocity observations
dating back to 1897. They presented two solutions. One uses the velocities outside the eclipse phases togethher
with mid-times of previous eclipses, from photometry dating back to 1842, which provide strongest constraint
on the ephemeris. From this they find a period of 9896 days (27.0938 years) and an orbital eccentricity
 of 0.227. By using only radial velocities
they find that the predicted middle of the current eclipse is nine months earlier, implying that the
gravitating companion is not the same as the eclipsing object. They conclude that, the purely spectroscopic
solution may be baised by perturbations in the velocities due to the short-period oscillations of the
F0Iae primary star. Other notable recent results are the infrared images of the transiting disk by
Kloppenborg et al. (2010) and interferometric studies by Stencel et al. (2008). More recently Stencel et al. (2011)
made detailed infrared studies of $\epsilon$ Aur during the 2009 - 2011 eclipse.\\

\noindent
Disks are among most common astrophysical systems, and a powerful technique for observing them is stellar
occultation method (Forrester $\&$ Lissauer 1982). $\epsilon$ Aurigae offers this opportunity to a new  class of
disks which are associated with post-AGB stars. $\epsilon$ Aur disk was described variously as thick (Huang 1965)
 or thin, flat or twisted,
opaque or semi transparent, fully solid or possessing a central hole. A wealth of data are now available in the
literature from UV to the far-IR wavelengths, which can give  an insight into the nature of the disk. \\

\noindent
From the SED constructed by their new $Spitzer$ $Space$ $Telescope$ observations and the archival far-UV to mid-IR
data, Hoard et al.(2010) proposed a three-component model for the $\epsilon$ Aur system which consists of a F0Iae
post-AGB star and a B5V type main sequence star surrounded by a geometrically thin and partially transparent
disk. They proposed a single-temperature black body model for the disk and constrained the disk to have a
temperature of 550K, a size of 3.8 AU with a thickness of 0.95 AU and is viewed nearly edge on. Their model deals
with average bulk properties of the disk with cylinderical volume and assumes an uniform mass distribution. However,
the nature of the disk with more realistic characteristics such as radial density distribution  and temperature
profile, scale height and grain chemistry obtained through a 2D radiative transfer modeling and addressing its origin
are not yet available. \\

\noindent
Near-IR interferometric imaging of $\epsilon$ Aur in the H band was made in November 2009 and in December 2009 by
Kloppenborg et al. (2010) which showed that the eclipsing body is a tilted opaque disk which is moving in front of
the F star. Their study reveals the compactness of the obscuring disk across the two epochs and provides the first
direct evidence for the presence of a geometrically thin and optically thick disk. With the estimated Hipparcos
distance of 625pc the maximum thickness of the disk obtained by them is 0.76AU and the radius of the disk is 3.81
AU. Kloppenborg et al.(2010) also estimated the mass of the F star to be 3.69M$_{\odot}$. They estimate a disk mass
of 4.45 $\times$ 10$^{-5}$M$_{\odot}$ (with ISM gas-dust ratio).\\

\noindent
Results obtained by Kloppenborg et al.(2010) from their interferometric imaging observations can greatly help to
make a detailed disk model of $\epsilon$ Aur describing its nature with density and temperature profiles and disk
grain chemistry possible. Here we attempt to make such a model for the disk from fitting of the specral energy
distribution (SED) of $\epsilon$ Aur from far-UV to far-IR wavelengths obtained from archival data and by solving
the radiative transfer problem of the disk in two dimensional case. \\

\noindent
\begin{center}
{\bf {2. SED of $\epsilon$ Aur}} \\
\end{center}
\noindent
To construct the SED of $\epsilon$ Aur from far-UV to the far-IR wavelengths we have used the archival data from ground based and
space based observing facilities. UV data were obtained from $HST$ GHRS, optical and near-IR
photometric data were taken from $SIMBAD$, and mid-IR to far-IR photometric measurements were obtained from $IRAS$ and $MSX$ space
missions. Mid-IR spectra from 9.89 to 37.14 $\mu$ were taken from $Spitzer$ $Heritage$ $Archive$. Photometric measurements of Herschel 
infrared bandpasses
were taken from recently published results of Hoard et al. (2012). We have also obtained new mid-IR and
far-IR fluxes at five photometric bands of AKARI, viz. at 9.0$\mu$ 18.0$\mu$, 65.0 $\mu$, 90.0$\mu$ and 140.0$\mu$, see Table 1. Wherever
magnitude measurements were available, they were converted into fluxes using appropriate zero magnitude fluxes at the respective
photometric bands. All the observations were made outside the eclipse phases of $\epsilon$ Aur. Majority of these observations were
obtained prior to the onset of the 2009 eclipse but well after the end of the 1984 eclipse.\\ 

\noindent
\begin{center} 
{\bf {3. Radiative Transfer Modeling}}\\
\end{center}

\noindent
To solve the radiative transfer problem in the disk of $\epsilon$ Aur in two dimensions, we have used a Monte-Carlo radiative transfer
code $SRCDUST$. It is based on the Monte-Carlo radiative equilibrium and temperature correction techniques developed by Bjorkman $\&$
Wood (2001) and was adopted to simulate ellipsoidal envelopes and T Tauri disks (Wood et al. 2002), Whitney et al. (2003). This
code was tested by comparing to a set of benchmark calculations for spherically symmetric codes by Bjorkman $\&$ Wood (2001). This code
can solve the radiative transfer problem in three dimensional cases and can well be applied to astrophysical systems having axial
symmetry geometry with disk, envelope and outflow components which are illuminated by a central star.  . Physical properties of the star
and physical and geometrical parameters of the disk with an appropriate dustmodel are provided as the input for the code.  We have
chosen the code to consider only the disk component illuminated by a star at the center. $SRCDUST$ provides disk temperature structure
and synthesized SED of the star and the disk at required angle of view  in the output. More details on the code can be seen in Whitney
et al. (2003). The disk is considered here as to have formed by accretion process and hence we use a standard flared accretion  density
structure (Lynden-Bell $\&$ Pringle 1974) described as \\

\noindent
$\rho$ = $\rho_{0}(1-(R_{star}/\omega)^{0.5}) (R_{star}/\omega)^{\alpha})exp{-1/2(z/h(\omega))^{2.0}}$ \hspace{1cm} Eqn. 1. \\
 
\noindent 
Where, $\omega$ is the radial coordinate of the disk midplane and the scale height increases with radius as $h =
h_{0}(\omega/R_{star})^{\beta}$. For our models we adopt flaring parameter $\beta$ = 1.25 based on the accretion  disk models at hydrostatic
equilibrium (D'Alessio et al. 1999) and was used to describe the structure of the accretion disk (Wood et al. 2001, Vinkivic 2012, Thi, Woitke
$\&$ Kamp 2011) and the value of $\alpha$ = 3($\beta$ - 0.5) = 2.25 (Shakura $\&$ Sunyaev 1973). We take $h_{0}$ = 0.05$R_{star}$ such
that the disk will have a thickness of 0.76AU at its outer edge of radius 3.8AU as constrained by the IR interferometric observations of
Kloppenborg et al. (2010). The inner radius is constrained by dust sublimation temperature. The angle of inclination and the mass of the disk
are varied to match the synthesized SED with the observations. For this specific case, the primary of $\epsilon$ Aur can also heat the disk,
and evidence for an increase in tempeature in the F star heated portions were observed in the IR spectra during the post mid-eclipse recently
by Stencel et al. (2011), and the effect of irradiation was studied by Takeuchi (2011). $SPITZER$ data of Hoard et al. (2010), obtained when
the orbital phase is 0.8, differs from MSX measurement obtained at phase 0.5 which are systematically brighter in the 3-5 $\mu$ region. 
Photometric flux at $IRAC$ 4.47$\mu$ band is 52.9 Jy and at 4.35 $\mu$ $MSX-B2$ band it is 72.1 Jy. This can be attributed to viewing the
hotter side of the disk irradited mostly by the F star (Hoard et al 2010, Taranova et al 2001).  The systematic difference suggests an actual
difference in the characteristics of the cool componant between these two observations and no time variability on this has been established.
However, disk heating by the primary is not considered in our models as overall disk heating is dominated by the hard UV  photons from the 
hot central star, and recently taken $SPITZER$ data where the differential heating is minimum is given more important for modeling. This is
adequate to study the disk physical and chemical properties. \\

\noindent 
\begin{center}
{\bf {4. Results}} \\
\end{center}

\noindent
We have computed model SEDs of the disk with the central star with different physical and chemical properties of dust
in the disk. The central star has a spectral type of B5V with $T_{eff}$ of 15,000 K and having mass of  5.9
$M_{\odot}$ and a surface gravity log g of 4.0 (Hoard et al. 2010). Models for the disk were computed with 1) grains
with the ISM size distribution and having amorphous carbon chemistry 2) grains with ISM chemical composition
following MRN size distribution but having sizes much larger than the ISM grains (10$\mu$ to 100 $\mu$) and  3)
grains with large grain sizes following MRN distribution function but having silicate alone and amorphous  carbon
alone dust chemistry. \\

\noindent
For a
given dust chemistry and for a given grain size distribution function, a dustmodel with absorption and scattering cross
sections, the mass absorption coefficient and the cosine asymmetry parameter (g factor) for a wavelength range of 0.005
$\mu$ to 900 $\mu$ is computed using a mie code. Grains are taken to be spherically symmetric and we have used
Henyey-Greenstein phase function (Henyey $\&$ Greenstein 1941) characterised by the g factor. This dustmodel file is
called as an input by the $SRCDUST$ radiative transfer code.\\

\noindent
The wavelength dependent optical constants for the astronomical silicates and amorphous carbon required to calculate the dustmodel
file were taken from Draine (2003) and Zubko et al. (1996) respectively for all our models. The outer radius of the disk R$_{out}$ is
adapted from Kloppenbourg et al. (2010), which is 3.8 AU and the inner radius R$_{in}$ is decided by the dust sublimination temperature
at the disk which is taken as 1500 K. Each model was taken from best out of about 10  models computed. The flux output from the
radiative transfer code is then added to the Kurucz model flux at each wavelength of a F0Iae post-AGB star with $T_{eff}$ = 7700, log g
$\sim$ 1 (Castelli, Hoekstra $\&$ Kondo, 1982) and mass of 2.2 $M_{\odot}$ with solar abundance (Castelli $\&$ Kurucz, 2003) to obtain
the final SED of $\epsilon$ Aur. The model SED derived by this method was then subjected to the interstellar extinction with a value of
A$_{V}$ = 1.1 found in the literature (Mozurkewich et al. 2003), which can be directly compared with the observations. The interstellar
extinction curve needed for calculating ISM contribution was computed  empirically for all wavelengths using the method given by
Fitzpatric $\&$ Massa (2007).  \\

\noindent
In the following sections we discuss individual models of the disk in detail and compare them against the 
mutiwavelength observations to constrain the nature of the grains in the disk and disk geometry and discuss the 
origin of the disk. \\

\noindent
\begin{center}
{\bf 4.1 Larger grains in the disk} \\
\end{center}
\noindent
As noted earlier by Kopal (1971) the eclipse of $\epsilon$ Aur in the optical and near-IR wavelengths was observed to
be approximately gray, indicating a larger grain population in the disk. Broad dust features are expected in the
spectra if the grain sizes are much smaller than the central wavelength of the  emission feature. When 2 $\pi$ a 
(a is the size of the grain) is
smaller than $\lambda$, constant emissivity will be seen for strongly absorbing materials and no spectral features
will be found. As it was seen, the far-IR spectra of $\epsilon$ Aur obtained with $SPITZER$ $IRS$ are smooth without
any notable dust features (Hoard et al. 2010), sugesting a grain sizes larger than 10 $\mu$. Lack of solid state
features were also noticed by Stencel et al. (2011) in their IR spectra. We examine here using radiative transfer
models the absence of such broad dust features in the spectra of $\epsilon$ Aur by taking large grain size and further
propose here the expected grain sizes in the disk of $\epsilon$ Aur. We have
calculated dust models in the disk following MRN grain size distribution and one having ISM grain sizes with
$a_{min}$  =  0.05$\mu$ and $a_{max}$ = 0.2$\mu$ (model 1) and another having larger grain sizes with $a_{min}$ =
10$\mu$ and a$_{max}$ = 100 $\mu$ (model 2). Both the dust models have amorphous carbon grain chemistry. These dust
models are used for our radiative transfer simulation. \\

\noindent
Monte-Carlo radiative transfer in the disk was calculated with 10 million photons from the central star, to minimize
ripples seen in the SED at longer wavelength region. Models with different inclination angles of the disk were
calculated, and the disk model corresponds to nearly edge-on viewing  fits the observations better than others. 
The model SED of the $\epsilon$ Aur system viewed at 87$^{o}$ and at 60$^{o}$ are shown against observations in Fig 2. 
The thermal images of the disk viewed at these two incliation angles and at three different wavebands (K band, $SPITZER$ IRAC 8$\mu$m 
and $SPITZER$ MIPS 70 $\mu$ bands) are shown in Fig 5. The grain sizes for model 2 were arrived by gradually
increasing the lower and upper limits of grain sizes form model 1. The minimum size is fixed at 10 $\mu$ by the disappearance of the
IR features and the maximum size could be even larger as no notable changes are seen at longer wavelength side of the considered range.
Sub-mm fluxes will say more on it.  \\

\noindent
The computed model SEDs using $SRCDUST$, after subjecting to the ISM extinction, were compared with the observations
in Fig 1. It can be seen that model with smaller grain sizes produce broad dust emission features in the far-IR
region of the SED and the overall match to the observation is not good. Whereas the model SED corresponding to the
large grain population is nearly smooth at all wavelengths. The plot of computed dust opaciy againt wavelength for the ISM grain
size distribution is also shown in the lower panel of Fig 1. As seen in the model 2, the disk is sufficietly hot to show 
the features in emission if the grains are small in sizes, the minimum temperature observed at the outer edge of the
disk is 252 K for the model 2.  \\

\noindent
The radial temperature profile at  the disk midplane is shown in
Fig. 3. We conclude from our study that the grains in the disk of $\epsilon$  Aur are much larger in size than 
the ISM grains. This may indicate a possibility of grain growth in the disk of $\epsilon$ Aur. Mass of the disk is found
to be low, 0.005 $M_{\odot}$ for the assumed gas to dust mass ratio of 100. The SEDs do not fit with the observations
in the shorter wavelength region (see Fig 2.), which may indicate an additional grain population with very small grains
and having a wavelength depended opacity at UV region as noted by Hoard et al. (2010). \\

\begin{center}
{\bf 4.2. Grain Chemistry and the origin of the disk}
\end{center}
\noindent
{4.2.1 Disk with ISM grain chemistry} \\

\noindent 
An important test which will give an idea on the origin of the disk is the grain chemistry in the disk. If the disk
is a protoplanetary disk then we expect the grain chemistry in the disk to be a mixture of amorphous silicates and
amorphous carbon grains with mass proportion as seen in the ISM. We have run a radiative transfer model to investigate
if the SED shows ISM dust chemistry. For the ISM dust model we have used bare spherical grains composed with 60$\%$
of astronomical silicate and 40$\%$ of amorphous carbon in mass. As we have seen that the grains in $\epsilon$ Aur disk
are larger in size, we have taken a grain size distribution with a$_{min}$ = 10.0$\mu$ and a$_{max}$ = 100$\mu$
following standard MRN power law with an exponent of q = -3.5 (model 3) to compute the dustmodel file required for
the radiative transfer simulation. \\

\noindent
Monte Carlo radiative transfer through the disk was calculated for this dustmodel with 10 million photons from the
central star to get the model SED. After subjecting to the ISM extinction, the best fit SED to the observation implies
that the disk is inclined to the line-of sight with an angle larger than 85 degrees. The disk mass was estimated to
be 0.005 $M_{\odot}$ assuming a gas to dust mass ratio of 100.  The computed SEDs corresponding to this model is
plotted against the observed data points in Fig 3. As seen in the figure, the model fit is not good enough. It shows
a shallower slope in the far-IR region when compared with the observations. This implies that the grain chemistry in the disk
of $\epsilon$ Aur does not resemble with the ISM grain chemistry. The radial temperature profile in the disk midplane
for this model is shown in Fig 3. The minimum temperature at the outer edge of the disk for this model is 292 K. \\

\noindent
{4.2.2  Disk with post-AGB envelope dust chemistry}  \\

\noindent 
To investigate if the disk was originated from the mass transfer and/or accretion from the post-AGB primary  on to
its secondary, we have computed SEDs for disk models with silicate alone and amorphous carbon alone grain chemistry.
If the primary were to be an evolved AGB star tranfering mass through the lagrangian point by Roche lobe overflow to
its companion, then the dust grains in the disk should have pure silicate or pure amorphous carbon chemistry, depending
on the mass and the evolutionary stage of the AGB primary at the time of mass transfer and/or accretion. We have made
two dustmodel files seperately with amorphous silicate and  amorphous carbon  grains, both following a MRN grain size
distribution with sizes varying from a$_{min}$ = 10$\mu$ and a$_{max}$ = 100 $\mu$ with a power law exponent of -3.5.
Monte Carlo radiation transfer through the disk was computed for these two dustmodels with 100 million photons from
the central star to get a smooth SED at longer wavelengths. The SED obtained from the simulation was subjected to the
interstellar extinction and then compared with the observations (see Fig.3).\\

\noindent
Our results imply that the disk model with amorphous carbon dust chemistry fits the observations better than the
amorphous silicate dust chemistry which shows a steeper slope in the far-IR region.  We suggest from our study that
the grains in the disk of $\epsilon$ Aur are basically amorphous carbon, and it is quite unlikely that silicate dust
is present in the disk. Hence we conclude that the disk was formed from  mass transfer and/or accretion from the C rich
post-AGB star (during superwind mass-loss phase on AGB) to its main sequence companion. We took a gas to dust mass ratio of 100 which  resulted a disk mass
less than 0.005 $M_{\odot}$. The radial  temperature structure of the disk corresponding to these two models are
shown in Fig. 4. The minimum grain temperature is seen at the outer edge of the disk has a value of  293 K for
amorphous silicate model (model 4) and 252 K for amorphous carbon model (model 2). \\

\begin{center}
{\bf 4.3 Does the disk have a central void ?}
\end{center}
\noindent
It was discussed earlier in the literature if the disk of $\epsilon$ Aur has a central void or not. Hoard et al.
(2010)  argued on the presence of a void at the center of the disk which was originally proposed by Wilson (1971) and
Wilson $\&$ Van Hamme (1986) to explain the  mid-eclipse brightening observed during 1982-1984 eclipse. Our radiative
transfer models also comply with this suggestion. For the given parameters of the central star and the disk, the
radial temperature structure of the disk computed by the code for all models are shown in Fig 4. As seen in the
figure, the dust sublimation temperature of 1500K, which determines the inner edge of the disk which is starting at
around 2.0 AU from the central star for all the models considered.  If the central star is hot, as observed in the UV
data (Parthasarathy $\&$ Lambert 1983b) a void near the star is expected and our study gives a quantitative estimate
for the size of this void. Within this radius of 2 AU the matter in the  disk is dust free and hence it is more
transparent. It is hence proposed that mass distribution in the disk follows Eq. 1, from the outer edge of the disk
(3.8 AU) to the inner edge (2.0 AU) and the relation breaks below this radius. It is possible that gaseous matter may 
be present in the void in nuetral or ionized form which is free from dust. This void could cause the mid eclipse
brightening. Fig 5 shows the thermal images of the disk simulated for IRAC and MIPS bands from our code for inclination
of 87$^{o}$ and 60$^{o}$. \\

\noindent
We propose that the clearing of dust in the central hole of $\epsilon$ Aur disk is due to the photo evaporation processes (Clarke
et al. 2001) as the inner edge of the disk is produced by the dust sublimation temperature. The transition region of the
inner edge of the disk to the central hole  is hence expected to be smooth, unlike for the case of the disk observed with
LkH$\alpha$330, where the dust clearing  in the hole could be caused by gravitational perturbation (Brown, Black, Qi et
al. 2008). The inner edge of $\epsilon$ Aur disk in our study is like the fixed structure model considered by Thi, Woitke $\&$ Kamp (2011)
which differs from their soft edge models with rounded inner rims showing enhanced near-IR continuum emission when
viewed face on. For the case of $\epsilon$ Aur, where there is no enhanced near-IR emission observed and the disk is 
viewed edge on, it is difficult to constrain the existence of a puffed up inner edge from its SED.

\begin{center}
{\bf 5. Conclusions and Discussion}
\end{center}

\noindent
From solving the radiative transfer problem of the disk of $\epsilon$ Aur in 2D case, we conclude that the disk is less massive
(0.005$M_{\odot}$) and is seen nearly edge on. The dust grains in the disk are much larger than the ISM grains having sizes of 10
$\mu$ to 100 $\mu$ and have carbonaceous dust chemistry. Silicate is not expected to be present. This shows a C rich post-AGB as the
primary of $\epsilon$ Aur. New AKARI data presented here fit well with the proposed model SED of $\epsilon$ Aur, however at 140 $\mu$
the deviation is significant, even after taking into account of the error in the measurement. This may show existence of more cool
dust in the outer rim on the disk which is not considered by the model. More data in the far-IR region (near 140 $\mu$) is needed to
confirm this. The temperature structure of the disk for the computed models shows that the disk has a central void of radius 2 AU.
This void causes a mid eclipsing brightening. Recently Stencel et al. (2011) from infrared studies of $\epsilon$ Aur during the
recent eclipse conclude that the disk is dominated by large grains. Bipoar dusty and gaseous disks have been detected around young and
evolved stars from followup high resolution imaging surveys of IRAS sources (Chesneau 2010). Amomg the post-AGB stars there are
several objects with bipolar dust disks and some of them were found to be single lined spectroscopic binaries  (Chesneau 2010).
Circumbinary dust disk has been found around the evolved binary Upsilon Sagittarii (Netolickey et al. 2009). Also, large H-alpha
formimg region has been found around Beta Lyrae and Upsilon Sagittarii (Bonneau et al. 2011). The H-alpha profile in the out side
eclipse spectra of $\epsilon$ Aur suggests presence of similar large H-alpha envelope or ring around the F0Iae star. Shell spectra
produced by the gaseous envelope of the disk, as presented by Ferluga $\&$ Mangiacapra (1991) can  constrain if the void is empty or
filled with nuetral and ionized gases. IR spectra will help as it is expected that  the IR spectrum will be dominated by the nebular
lines of the disk than the photosphere of the primary (Stencel 2007). The disk mass estimated from our study can be taken only as the
lower limit as it is the mass of an annular disk with inner radius 2 AU and outer radius 3.8 AU. \\

\begin{center}
{\bf 6. Acknowledgements}
\end{center}
We are thankful to Prof. Yoshifusa Ita for providing us the AKARI infrared fluxes of Epsilon Aurigae.
MP is thankful to Prof. Shoken Miyama and Prof. Yoichi  Takeda
  for their kind support, encouragement and hospitality. We are thankful to the referee for valuable comments which have improved
  the paper significantly.  \\
	       
\noindent
\begin{center}

{\bf 6. References}\\
\end{center}
\noindent
Ake,T.B. $\&$ Simon, T., 1984, Future of UV Astronomy Based on Six Years of IUE Research, J.M. Mead et al. eds., p. 361 \\
Altner, B., et al. 1986, A $\&$ A Suppl. 65, 199 \\
Bjorkman, J.E. $\&$ Wood, K. 2001, ApJ, 554, 615 \\
Brown, J.M., Blake, G.A., Qi, C., et al. 2008, ApJ., 675, L 112 \\
Boehm.C., Ferluga, S. $\&$ Hack, M., 1984, A $\&$ A 130, 419 \\
Bonneau, D., et al. 2011, A $\&$ A 532, A148 \\
Castelli, F., Houkstra, R. $\&$ Kondo, Y. 1982, A$\&$AS, 50, 233 \\
Castelli, F. $\&$ Kurucz, R.L. 2003, in IAU Symp 210, Modelling of stellar Atmospheres, ed. N. Piskunov,W.W
Chapman, R.D., Kondo,Y., $\&$ Stencel, R.E., 1983, ApJ 269, L17 \\
Clarke, C.J., Gendrin, A., $\&$ Sotomayor, M. 2001, MNRAS, 328, 485 \\
D' Alessio, P., Calvet, N., Hartmann, N. et al. 1999,ApJ, 527, 893 \\
Weiss, $\&$ D.F. Gray (San Francisco, CA. ASP), Poster A20 \\
Carroll, S.M.,Guinan, E.F., McCook, G.P. et al. 1991, ApJ 367, 278 \\
Draine, B. 2003, ApJ, 598, 1017 \\
Ferluga,S. $\&$ Hack, M. 1985, A$\&$A 144, 395 \\
Ferluga,S. $\&$ Mangiacapra,D. 1991,  A $\&$ A 243, 230 \\
Fitzpatrick, E.L. $\&$ Massa D. 2007, ApJ, 663, 320 \\
Guinan, E.F. $\&$ Dewarf,L.E. 2002 in Exotic stars as challenges to evolution,  ASP Conf series 279, 121 \\
Hack, M. 1961, Mem.Soc.Astron. Ital, 32, 351 \\
Henyey, L.G. $\&$ Greenstein, J.L.,  1941, ApJ, 93, 70 \\
Hoard, D.W., Howell, S.B $\&$ Stencel, R.E. 2010, ApJ, 714\\
Hoard, D.W., Ladjal, D, Stencel, R.E. et al. 2012, ApJL \\
Huang, S., 1965, ApJ, 141, 976\\
Kloppenborg, B., Stencel, R.E. $\&$ Monnier, J. et al. 2010, Nature, 464, 870 \\
Lynden-Bell, D. $\&$ Pringle, J.E. 1974, MNRAS, 168, 603 \\
Mozurkewich,D. et al., 2003, AJ, 126, 2502 \\
Netolicky, M., et al. 2009, A$\&$A, 499, 827 \\
Parthasarathy,M., 1982, in Epsilon Aurige News Letter,  No. 5,  page 5, December 1982, edited by R.E. Stencel \\
Parthasarathy, M. $\&$  Frueh, M.L. ApSS, 123, 31 \\
Parthasarathy,M. $\&$ Lambert, D.L., 1983a, IAU circular 3766 \\
Parthasarathy, M., $\&$ Lambert, D.L., 1983b, PASP 95, 1012 \\
Perryman, M.A.C. et al. 1997, A$\&$A 323, L49 \\
Sadakane, K., Kambe, E., Sato, B. et al. 2010, PASJ, 62, 1381 \\
Shakura, N.I. $\&$ Sunyaev, R.A. 1973, A$\&$A, 24, 337 \\
Stefanik, R., et al. 2010, AJ, 139, 1254 \\
Stencel, R. E., 2007, in binary stars as critical tools and tests in contemporary astrophysics
 IAU Symposium 240, 202 \\
Stencel, R.E., 1982, in Epsilon Aurigae News Letter, No. 5, page 5, December 1982, edited by R.E. Stencel \\
Stencel R. E, et al. 2008, ApJ, 689, L137 \\
Stencel, R.E., Kloppenborg, B.K., $\&$ Wall R.E., 2011, AJ, 142, 174 \\
Takeuchi, M. 2011, PASJ, 63, 325 \\
Thi, W.-F., Woitke, P.$\&$ Kamp, I. 2011, MNRAS, 412, 711 \\
Vinkovic, D. 2012, MNRAS, 420, 1541 \\ 
Whitney,B., Wood, K., Bjorkman, J.E. et al. 2003, ApJ, 591, 1049 \\
Webbink, R.F 1985, in NASA Cof. Publ. 2384, North American Workshop on Recent Eclipse of Epsilon Aurigae, ed. R.E.Stencel
(Washington, DC: NASA). 49 \\
Wilson, R.E., 1971, ApJ, 170, 529 \\
Wilson, R.E., $\&$ Van Hamme, W., 1986, in IAU Hightlights of Astronomy, 7, 205\\
Wood, K., Wolf, M.J., Bjorkman, J.E $\&$ Whitney, B. 2002, ApJ, 564, 887 \\
Woolf, N.J., 1973, ApJ, 185, 229\\
Zubko V. et al 1996, MNRAS, 282, 1321-1329 \\

\noindent

\newpage

\vspace{2.2cm} \protect \begin{figure}[hp]
\vspace{4.cm}
\includegraphics{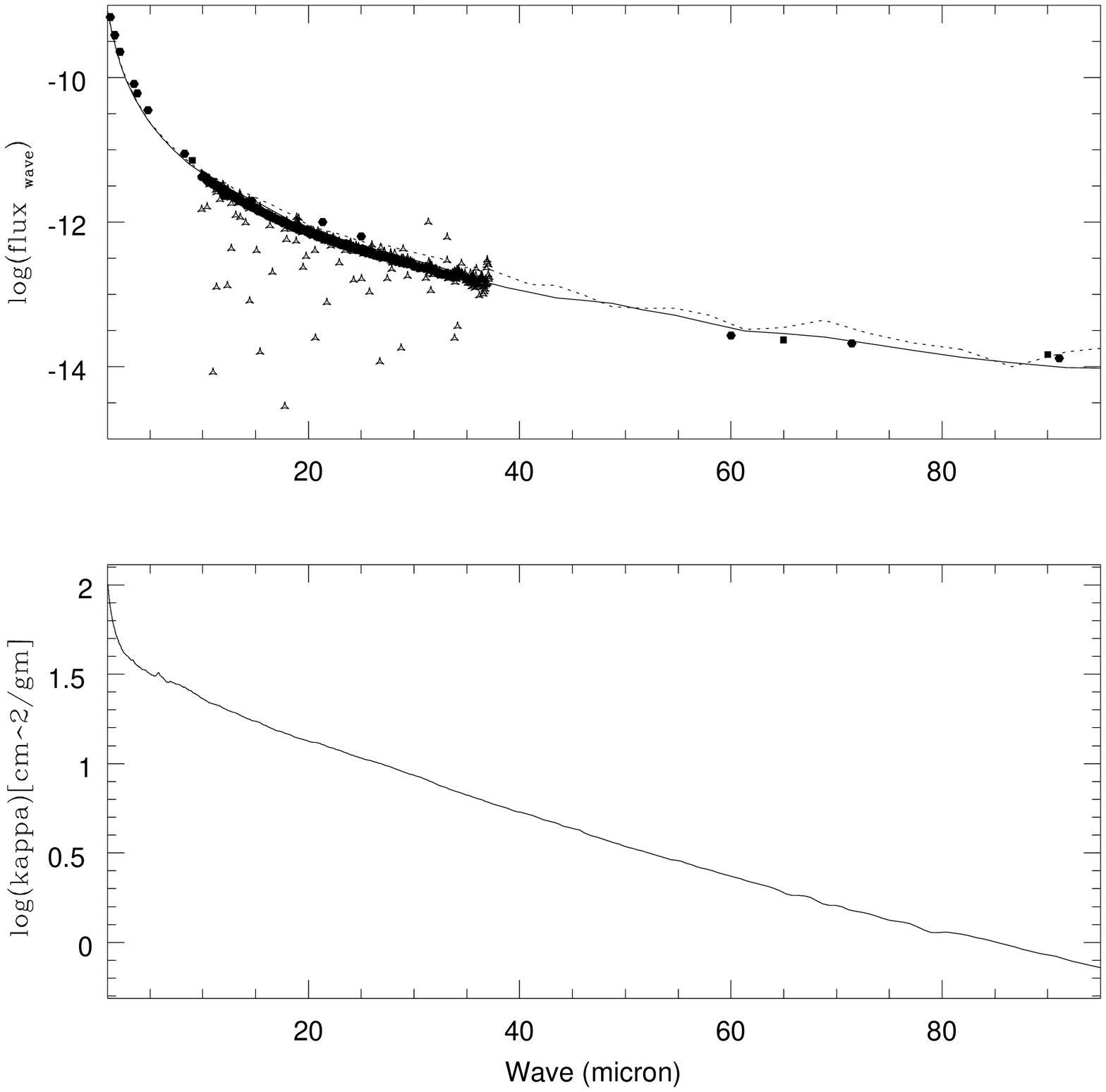}
\vspace{8.3cm}
\caption{Top: Model SEDs of the disk with large size carbonaceous grains (solid line) and carbonaceous
grains with ISM grain size distribution (dashed line). New AKARI fluxes are indicated in filled squares, Herschel
measurements are shown in square with caps, SPITZER IRS spectra
is shown in filled triangles and other photometric measurements are in filled pentagon; Bottom: mass absorption coefficient of
amorphous carbon (for unit gram of gas and dust) with ISM grain size distribution calculated using optical constants taken from
 Zubko et al.(1996). } 
\end{figure}

\newpage
\vspace{2.2cm} \protect \begin{figure}[hp]
\vspace{4.0cm}
\includegraphics{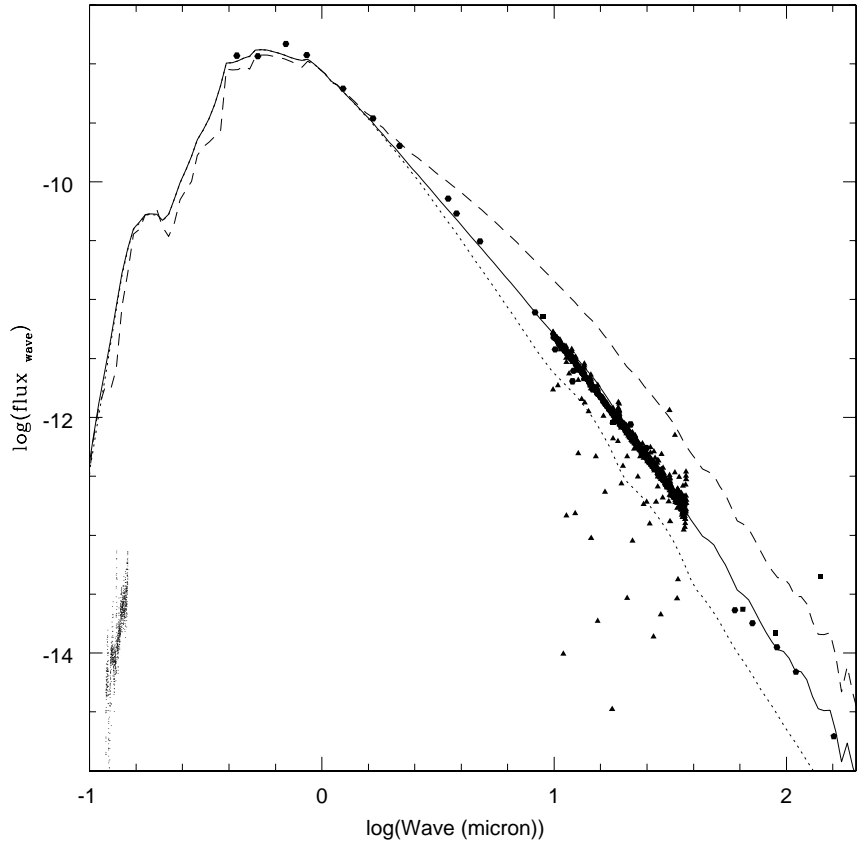}
\vspace{8cm}
\caption{ Model SED fit to the observations from UV to far-IR wavenengths; solid line shows SED of disk 
at an inclination of 87$^{o}$(model 2) and dashed line shows SED at 60$^{o}$ inclination.
GHRS data are shown in circular points and data description for other measurements are as in Fig 1. Dotted line indicate the stellar
photospheric flux without IR excess.}
\end{figure}

\newpage 
\vspace{2.2cm} \protect \begin{figure}[hp]
\vspace{4.cm}
\includegraphics{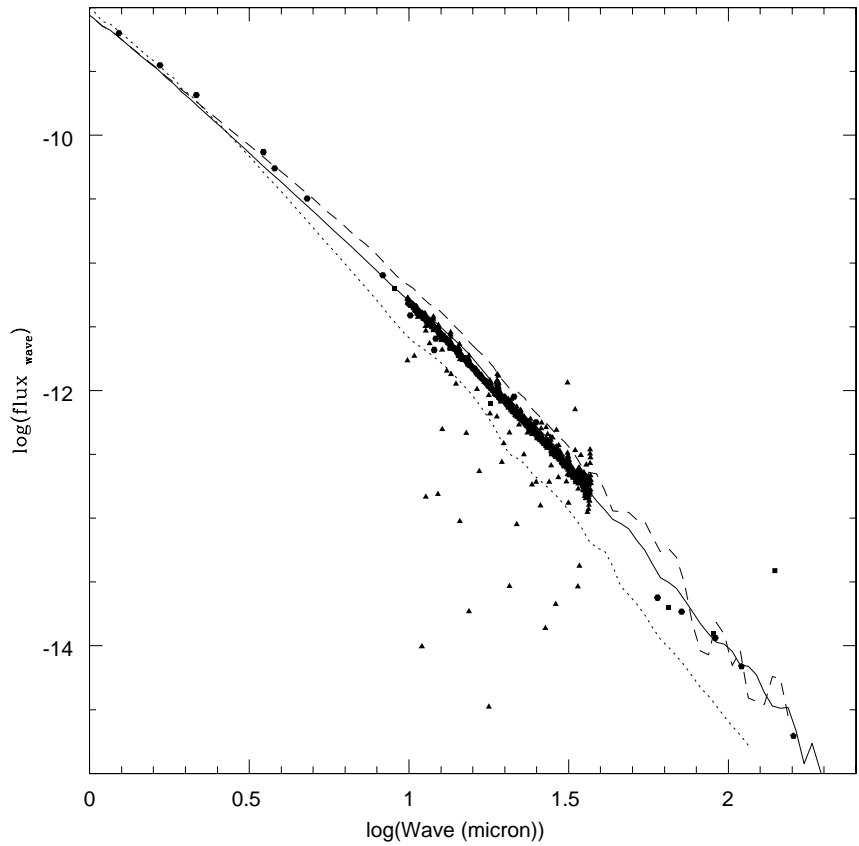}
\vspace{8.1cm}
\caption { Model SEDs of the disk with a) amorphous carbon grains (solid line) b) amorphous
silicate grains (dashed line) c) grains having ISM dust chemistry (dotted line). Data description is as in Fig 2.} 
\end{figure}

\newpage
\vspace{2.1cm} \protect \begin{figure}[hp]
\vspace{4.0cm}
\includegraphics{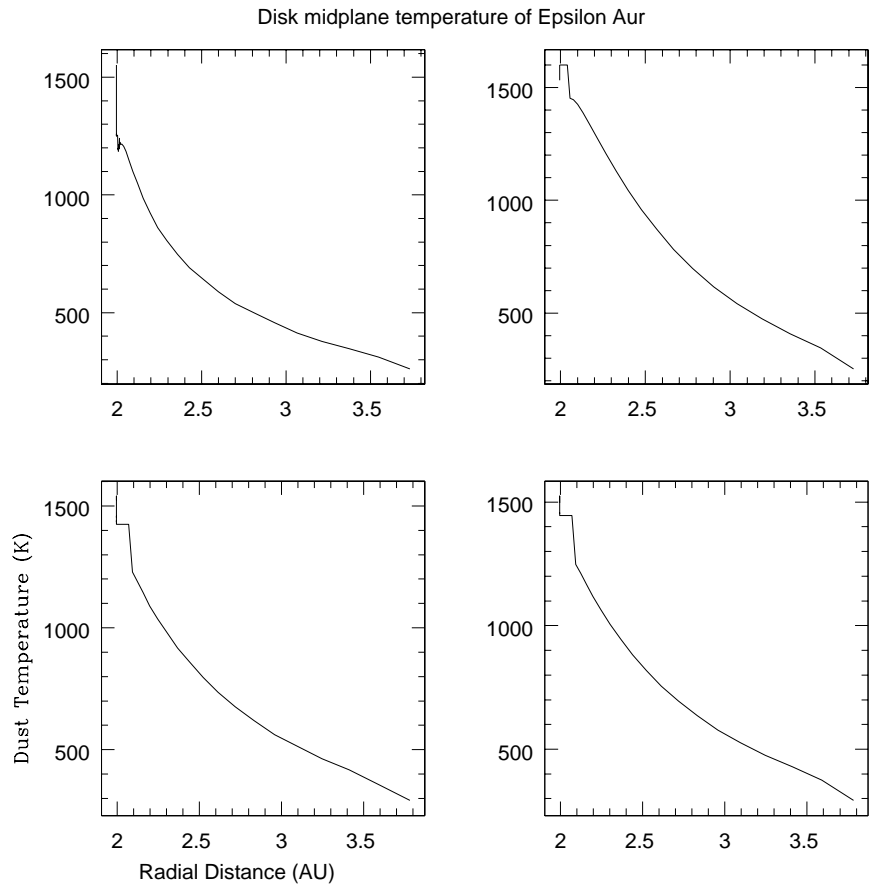}
\vspace{8.1cm}
\caption { Radial dust temperature profile of disk midplane. Model 1 (top left), model 2 (top right),
model 3 (bottom left) and model 4 (bottom right).}  
\end{figure}

\newpage
\vspace{2.1cm} \protect \begin{figure}[hp]
\vspace{1.0cm}
\includegraphics{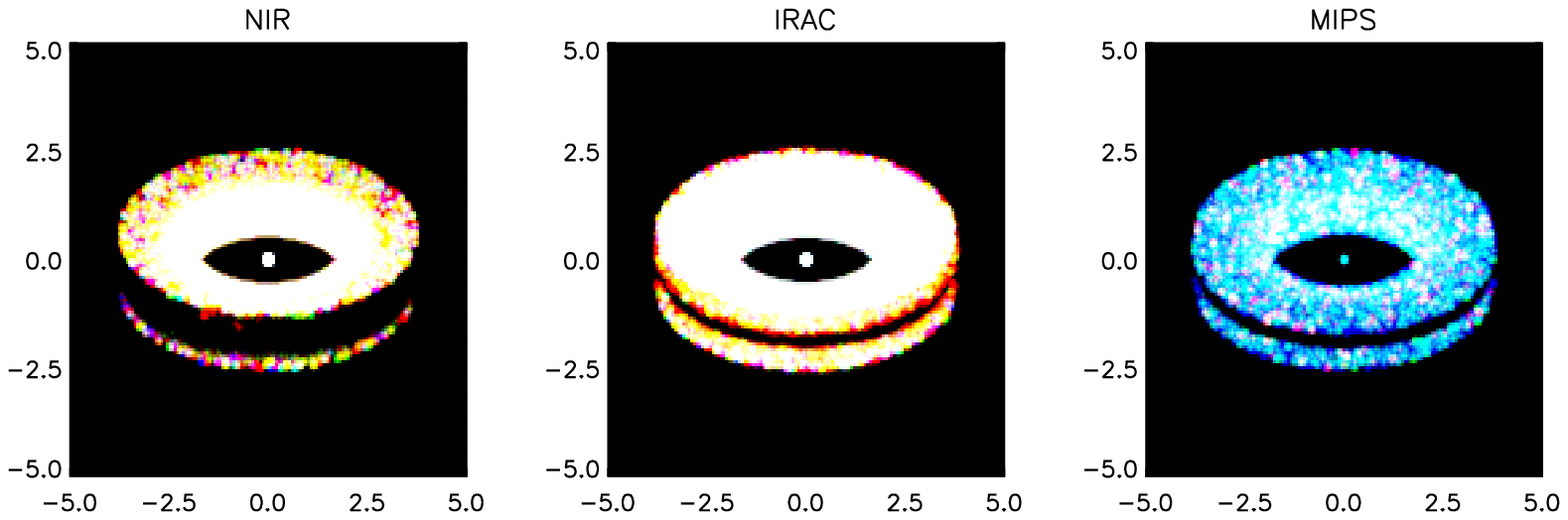}
\vspace{6.0cm}
\includegraphics{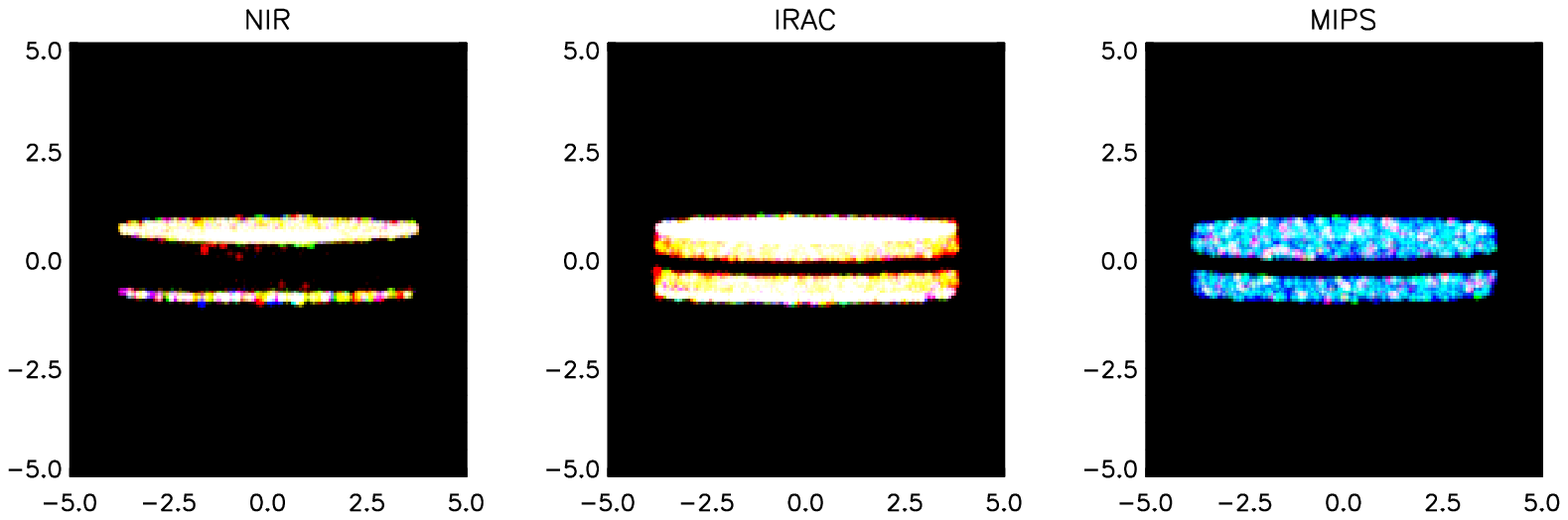}
\vspace{9.1cm}
\caption {Thermal images of $\epsilon$ Aur disk at K band, IRAC 8$\mu$  and MIPS 70$\mu$ bands obtained from our models 
at inclination angles 60$^{o}$ (top, showing the central void) and 87$^{o}$ (bottom).}  
\end{figure}

\newpage  
\newpage
\begin{tabular}{|c|c|c|c|}
\hline
Wavelength   & Flux & Flux quality &   Error\\      
 ($\mu$)       & (Jy)   &              & (Jy)   \\   
\hline
          &              &         &    \\
9.0       &   21.4143  & 3 & 0.128868     \\
          &           &   &               \\
18.0       & 5.44896  &  3 &  0.0576079  \\
           &          &    &             \\
65.0     & 0.509413   &   1  &  --        \\
         &            &      &             \\
90.0     & 0.44311    &  3   &  0.0615824  \\
         &            &      &              \\
140.0      &  2.11441   &   1  &  1.01968   \\
         &            &      &              \\  
\hline  
\end{tabular} \\

{Table 1: New AKARI data obtained in its  mid and far-IR bands.}
 \\

\begin{tabular}{|c|c|c|c|c|}
\hline
Models    & Grain Chemistry  & Grain Size distribution &   T$_{dust}$ at R$_{out}$ \\         
\hline
          &                        &                       &    \\
Model 1   & amorp. carbon &  $a_{min}$ = 0.05$\mu$ &   261 K \\
          &  &    $a_{max}$ =0.2 $\mu$ &   \\
          &                    &                    &           \\
Model 2   & amorp. carbon         &  $a_{min}$ = 10$\mu$ & 252 K     \\
          &                    &  $a_{max}$ = 100$\mu$&        \\
          &                    &                    &          \\
Model 3   & amorp. silicate 60$\%$& $a_{min}$ = 10$\mu$ &    292 K \\
          & amorp. carbon 40$\%$  &  $a_{max}$ = 100$\mu$ &   \\
          &                    &                    &          \\
Model 4   & amorp. silicate      &  $a_{min}$ = 10$\mu$ &  293 K   \\
          &                    &  $a_{max}$ = 100$\mu$ &       \\
          &                    &                    &          \\
\hline 
\end{tabular} \\
\vspace{3.cm}

{Table 2: Parameters of the radiative transfer models discussed in the text.}

\end{document}